\documentstyle[12pt]{article}

\newcommand{\sect}[1]{\setcounter{equation}{0}\section{#1}}

\newcommand{\eq}{\begin{equation}}
\newcommand{\eqa}{\begin{eqnarray}}
\newcommand{\en}{\end{equation}}
\newcommand{\ena}{\end{eqnarray}}
\newcommand{\enn}{\nonumber \end{equation}}

\def\sk{\vskip .4cm}
\def\noi{\noindent}

\def\al{\alpha}

\def\linv{{1 \over \lambda}}

\def\epsi{\varepsilon}

\def\de{\delta}

\def\part{\partial}

\def\R#1#2{ R^{#1}_{~~~#2} }
\def\PA#1#2{ P^{#1}_{A~~#2} }

\def\Rp#1#2{ (R^+)^{#1}_{~~~#2} }

\def\Rm#1#2{ (R^-)^{#1}_{~~~#2} }
\def\Rinv#1#2{ (R^{-1})^{#1}_{~~~#2} }

\def\Rpm#1#2{(R^{\pm})^{#1}_{~~~#2} }

\def\Rh{{\hat R}}

\def\MM#1#2#3{M^{#1~~~\!\!\!#3}_{~#2}}
\def\cchi#1#2{\chi^{#1}_{~#2}}

\def\U#1#2#3#4#5#6#7#8{U^{~#2~#4}_{#1~#3}|^{#5~#7}_{~#6~#8}}

\def\bu{\bullet}
\def\ci{\circ}

\def\T#1#2{ T^{#1}_{~~#2} }
\def\caM{{t}}
\def\M#1#2{ t^{#1}_{~\,#2} }

\def\rminus{r^{-1}}

\def\D{\Delta}

\def\Mat#1#2#3#4#5#6#7#8#9{\left( \matrix{
     #1 & #2 & #3 \cr
     #4 & #5 & #6 \cr
     #7 & #8 & #9 \cr
   }\right) }

\def\ep{\epsi^{\prime}}
\def\kp{\kappa^{\prime}}

\def\SqrNt{S_{q,r}(N+2)}

\def\Tc{{\cal T}}

\def\Lpm#1#2{L^{\pm #1}_{~~~#2}}

\def\LLpm{L^{\pm}}

\def\LLp{L^{+}}
\def\LLm{L^{-}}
\def\Lp#1#2{L^{+ #1}_{~~~#2}}
\def\Lm#1#2{L^{- #1}_{~~~#2}}
\def\Lc{{{\cal L}^+}}

\def\n2{{{N+1} \over 2}}
\def\ap{a^{\prime}}
\def\bp{b^{\prime}}

\def\square{{\,\lower0.9pt\vbox{\hrule \hbox{\vrule height 0.2 cm
\hskip 0.2 cm \vrule height 0.2 cm}\hrule}\,}}

\def\SO{SO_{q,r}(N+2)}
\def\ISO{ISO_{q,r}(N)}
\def\U{U_{q,r}(so(N+2))}
\def\IU{U_{q,r}(iso(N))}

\def\SqrNt{S_{q,r}(N+2)}

\def\H{H^\bot}
\def\le{\langle}
\def\re{\rangle}

\def\limrone{\lim_{r \rightarrow 1}}
\def\linv{{1 \over \lambda}}
\def\sma#1{\mbox{\footnotesize #1}}
\def\cvd{{\vskip -0.49cm\rightline{$\Box\!\Box\!\Box$}}\sk}

\begin{document}

\begin{titlepage}
\rightline{LBNL-39775}
\rightline{May 1997}
\vskip 4em
\begin{center}{\bf On the Geometry of the Quantum Poincar\'e Group}
\\[4em]
Paolo Aschieri\\[2em]
Theoretical Physics Group, Physics Division\\
Lawrence Berkeley National Laboratory, 1 Cyclotron Road \\
Berkeley, California 94720, USA.\\
\end{center}
\vskip 4em

\begin{abstract}
We review the construction of the multiparametric 
inhomogeneous orthogonal quantum group
$\ISO$  as a projection from $\SO$, and recall the conjugation that
for $N=4$  leads to the quantum Poincar\'e group.
We study the properties of the universal 
enveloping algebra $U_{q,r}(iso(N))$, and give an $R$-matrix formulation.
A quantum Lie algebra and a bicovariant differential calculus
on twisted $ISO(N)$ are found.
\end{abstract}

\vskip 8em

\noi{\small e-mail: aschieri@theorm.lbl.gov}~~~~~~~~~~~~~~~~~~~~~~~~~~~~
{}~~~~~~~~~~~~~~~~~~~~~~~~~~~~~~~~hep-th/9705227
\vskip .2cm
\noi \hrule
\vskip.2cm
\noi {\small {$~~~$}This work is supported  by a joint
                fellowship
                University of California --  Scuola Normale Superiore, 
                                                           Pisa, Italy.
It has been accomplished through the Director, Office of Energy Research,
Office af High Energy  and Nuclear Physics, 
Division of High Energy Physics
of the U.S. Department of Energy under Contract DE-AC03-76SF00098.}

\end{titlepage}
\newpage
\setcounter{page}{1}

\section{Introduction}

By quantum group, or noncommutative deformation of a Lie  group $G$,
we understand a
deformation of the algebra $Fun(G)$ of functions 
from $G$ to the complex numbers ${\mbox{\boldmath $C$}}$. 
$Fun(G)$ besides being a commutative algebra is a Hopf algebra, 
i.e. it has three 
linear maps called coproduct $\D~:~Fun(G)\rightarrow Fun(G)\otimes 
Fun(G)$,
counit $\epsi~:~Fun(G)\rightarrow {\mbox{\boldmath $C$}}$
and antipode (or coinverse) $\kappa~:~Fun(G)\rightarrow Fun(G)$,
defined by $\D f\,(g\otimes g')=f(gg')$, 
$\epsi(f)=f(1_G)$  and $(\kappa f)(g)=f(g^{-1})$.
These maps carry at the $Fun(G)$ level
the information about the product in $G$, the existence of the
neutral element $1_G$ and the existence of the inverse $g^{-1}$ of any 
$g\in G$. The information contained in the Hopf algebra $Fun(G)$ is the 
same as that contained in $G$, and we can work with $Fun(G)$
instead of $G$.

A quantum group is a deformation $Fun_q(G)$ of the algebra $Fun(G)$
depending on one or more complex parameters $q$. $Fun_q(G)$
has the same rich structure of $Fun(G)$ but it is no more commutative; 
the noncommutativity is given by the parameters $q$, when $q\rightarrow 
1$,
$Fun_q(G)\rightarrow Fun(G)$. 
\sk
Some motivations for studying $q$-groups in field theory 
have already been addressed 
in this conference by L. Castellani \cite{lproc};
our main concern here is that a noncommutative space-time,
with a deformed Poincar\'{e} symmetry group, 
is an interesting geometric background for the study of 
$q$-Minkowski space-time 
physics and,
in particular, of  Einstein-Cartan gravity theories 
based on the differential calculus on a $q$-Poincar\'e group
\cite{noncom}, \cite{Cas2}.
It is in this perspective that here, mainly reviewing 
\cite{altroarticolo} and 
\cite{ISODUAL}, we investigate
inhomogeneous orthogonal quantum groups, 
their quantum universal enveloping algebra, their quantum Lie algebras 
and more generally their differential structure. 
\sk
Various deformations of the Poincar\'e group are known in the literature;
here we will consider the quantum inhomogeneous orthogonal group $\ISO$ 
[and in particular 
$ISO_{q,r}(3,1; {\mbox{\boldmath $R$}})$]  found  as a projection from 
$\SO$. This projection 
procedure is then exploited to 
obtain $\IU$, the universal enveloping algebra of $\ISO$, 
as a particular Hopf subalgebra of $\U$;
an easy $R$-matrix formulation of $\IU$ is also given.
These results are accomplished through a
detailed study of the $\LLpm$ generators of  $U_{q,r}(so(N+2))$,
that are the basic elements needed to construct $q$-Lie algebras
and differential calculi.
Using these techniques in Section 6 we find a  $q$-Lie algebra and 
a differential calculus on twisted 
$ISO(N)$, i.e. the minimal deformation $ISO_{q,r=1}(N)$, obtained
with the parametric restriction $r=1$. 

To conclude
we mention that the differential geometry
on $\ISO$
naturally induces a differential structure on the $N$-dimensional 
orthogonal quantum plane, that can be seen as the $q$-coset space 
$\ISO/SO_{q,r}(N)$.
We thus have a canonical and straightforward way to obtain the
differential calculus on the 
$q$-Minkowski plane.

\section{$SO_{q,r}(N)$ multiparametric quantum group}

The $SO_{q,r}(N)$ multiparametric
quantum group is freely generated by the noncommuting
matrix elements $\T{a}{b}$ (fundamental representation 
$a,b= 1,\ldots N$) and
the unit element $I$, modulo the relation $\mbox{det}_{q,r}T=I$ and the 
quadratic $RTT$ and $CTT$ (othogonality) 
relations discussed below. The noncommutativity is controlled by the $R$  
matrix: 
\eq
\R{ab}{ef} \T{e}{c} \T{f}{d} = \T{b}{f} \T{a}{e} \R{ef}{cd}
\label{RTT}
\en
which satisfies the quantum Yang-Baxter equation
\eq
R_{12}R_{13}R_{23}=R_{23}R_{13}R_{12}~.
\label{QYB}
\en
The $R$-matrix components  
$\R{ab}{cd}$
depend
continuously on a (in general complex)
set of  parameters $q_{ab},r$.
{}For $q_{ab}= r$ we
recover the uniparametric orthogonal group $SO_r(N)$ of ref. \cite{FRT}. 
Then $q_{ab} \rightarrow 1, r \rightarrow 1$ is the classical limit for
which
$\R{ab}{cd} \rightarrow \de^a_c \de^b_d$ : the
matrix entries $\T{a}{b}$ commute and
become the usual entries of the fundamental representation. 
The $R$ matrix  is upper triangular 
(i.e. $\R{ab}{cd}=0$ if  [$\sma{$a=c$}$ and $\sma{$b<d$}$] or  
$\sma{$a<c$}$),
and the parameters $q_{ab}$ appear only in the diagonal components 
of $R$:
$\R{ab}{ab}={r/q_{ab}}, a \not= b,  \ap \not= b$,
where prime indices are defined as
$\ap \equiv N+1-a$. We also define $q_{aa}=q_{aa'}=r$.
The following relations reduce the number of independent
$q_{ab}$ parameters \cite{Schirrmacher}:
$      
q_{ba}={r^2/q_{ab}},~~
q_{ab}={r^2/q_{a\bp}}={r^2/q_{\ap b}}=q_{\ap\bp}~;
$
therefore
the $q_{ab}$ with $a < b \leq {N\over 2}$ give all the $q$'s.
\sk
Orthogonality  conditions are
imposed on the elements $\T{a}{b}$, consistently
with  the $RTT$ relations (\ref{RTT}):
\eq
C^{bc} \T{a}{b}  \T{d}{c}= C^{ad} I\;,~
C_{ac} \T{a}{b}  \T{c}{d}=C_{bd} I \label{Torthogonality}~,
\en
\noi where the matrix $C_{ab}$ 
and its inverse $C^{ab}$,
that satisfies $C^{ab} C_{bc}=\de^a_c=C_{cb} C^{ba}$, are the metric 
and its inverse. These matrices are antidiagonal;
they are equal, and for example for $N=4$ they read
$C_{14}=r^{-1}, C_{23}=1, C_{32}=1$ and $C_{41}=r$.
The explicit expression of $R$ and $C$
is given in \cite{Schirrmacher}, 
(see \cite{inson} for our notational conventions).
\sk
The coproduct $\D$, the counit $\epsi$ and the coinverse or 
antipode $\kappa$ 
are given by
\eqa
& & \D(\T{a}{b})=\T{a}{b} \otimes \T{b}{c}~,  \label{cos1} \\
& & \epsi (\T{a}{b})=\delta^a_b~,\\
& & \kappa(\T{a}{b})=C^{ac} \T{d}{c} C_{db}
\label{cos2}~.
\ena
One can also define  a $q$-antisymmetric
epsilon tensor and a quantum determinant $\mbox{det}_{q,r}T$
\cite{Fiore,ISODUAL};
to obtain the special 
orthogonal quantum group $SO_{q,r}(N)$ we impose also the
relation   $\mbox{det}_{q,r}T=I$. 
\sk
The conjugation that from the complex $SO_{q,r}(N)$
leads to the real
form $SO_{q,r}(n+1,n-1; {\mbox{\boldmath $R$}})$ 
and  that is in fact  needed to
obtain the quantum Poincar\'e group,
is defined by
$(T^a{}_b)^*={\cal{D}}^a{}_cT^c{}_d{\cal{D}}^d{}_b$,
${\cal{D}}$ being the matrix that exchanges the index $n$ 
with the index $n+1$ \cite{Firenze1}. 
This conjugation requires the following 
restrictions on the deformation parameters:
$|q_{ab}|=|r|=1$ for $a$ and $b$ both different from $n$ or $n+1$;
$q_{ab}/r\in {\mbox{\boldmath $R$}}$ when at least one of the 
indices $a, b$ is equal to  $n$ or $n+1$.
\sk
\section{$\ISO$ as a projection from \mbox{$\SO$}}
A fruitful way to introduce the $\ISO$ quantum group is to 
express it as a quotient of the Hopf algebra $\SO$. 
Let
$\T{A}{B}$ be the
$\SO$ generators, and  
split the index {\small A} of $\SO$  as
{\small A}=$(\circ, a, \bullet)$, with $a=1,...N$. 
With this notation:
\eq
\ISO= {\SO\over H}~,
\en
where $H$ is the left and right ideal in $\SO$  generated by 
the relations:
\eq
\T{a}{\circ}=\T{\bullet}{b}=\T{\bullet}{\circ}=0~.\label{Tprojected}
\en
Following \cite{inson}
the projection 
$P~:~\SO\rightarrow\SO/H$ is an epimorphism between
Hopf algebras,
and defining the projected matrix elements $\M{A}{B}=P(\T{A}{B})$,
we can give an $R$-matrix formulation of $\ISO$. We set
$u\equiv P(\T{\ci}{\ci}),~ y_b\equiv P(\T{\ci}{b}), $ 
$z\equiv P(\T{\ci}{\bu}),$ $x^a\equiv P(\T{a}{\bu})$
and (with abuse of notation)
$\T{a}{b}\equiv P(\T{a}{b})$, then we have
\sk
{\sl Theorem 3.1} The quantum group $\ISO$ is generated by the 
matrix entries
\eq
{\caM}
\equiv\Mat{u}{y_b}{z}{}{\T{a}{b}}{x^a}{}{}{v}
~~\mbox{ and the unity }~I
\en
modulo the $R\caM\caM$  and $C\caM\caM$ relations
\eq
\R{AB}{EF} \caM^{E}{}_{C} \caM^{F}{}_{D} = \caM^{B}{}_{F} \caM^{A}{}_{E} 
\R{EF}{CD}~,
\label{RTTISO}
\en
\eq
C^{BC} \caM^{A}{}_{B}  \caM^{D}{}_{C}= C^{AD} ,~ C_{AC} \caM^{A}{}_{B}  
\caM^{C}{}_{D}=C_{BD} 
\label{CMMbig}~,
\en
where $R$ and $C$ are the multiparametric $R$-matrix and metric of
$\SO$, respectively.

The co-structures are given by :
\eqa
& &\D (t^{A}{}_{B})=t^{A}{}_{C} \otimes t^{C}{}_{B}~;~~~
\nonumber\\
& &\epsi(t^A{}_B)=\delta^A{}_B~;\nonumber\\
& &\kappa(t^A{}_B)=C^{AC} t^{D}{}_{C} C_{DB}~.
\nonumber
\ena
\cvd
Using the explicit expression of the $R$ matrix one can check that 
relations
(\ref{RTTISO}), (\ref{CMMbig}) contain in particular the $SO_{q,r}(N)$ 
relations
(\ref{RTT}), 
(\ref{Torthogonality})
and the quantum orthogonal plane commutation relations:
\eq
\PA{ab}{cd} x^c x^d=0\label{qplane}
\en
where the $q$-antisymmetrizer $P_A$ is given by
\eq
P_A={1 \over {r+\rminus}} [-\Rh+rI-(r- r^{1-N})P_0] 
\label{proiett}
\en
and ${P_0}^{ab}_{~cd}\equiv(C_{ef}C^{ef})^{-1}C^{ab}C_{cd}$.
Moreover, due to the $Ctt$ relations, 
the $y$ and $z$ 
elements are polynomials in $u, x$ and $T$, and we also have
$uv=vu=I$.
A set of 
independent generators of $ISO$ is then 
\eq
\T{a}{b} , x^a , u , v\equiv u^{-1}
\mbox{ and the identity }  I~.
\en
Their commutations are (\ref{RTT}), (\ref{qplane}) and
\eqa
\!\!\!\!\!\!\!\!\!\!&&\T{b}{d} x^a={r \over q_{d\bullet}} 
\R{ab}{ef} x^e \T{f}{d}~,~\label{315a}\\
\!\!\!\!\!\!\!\!\!\!&&u \T{b}{d}=
{q_{b\bullet}\over q_{d\bullet}} \T{b}{d} u~,\label{uT}\\
\!\!\!\!\!\!\!\!\!\!&&u x^b=q_{b\bullet} x^bu~.\label{ux}
\ena
The deformation parameters of $\ISO$ are the same as those
of $\SO$; they are $r$ and $q_{AB}$ i.e. $r$, $q_{ab}$ and $q_{a\bu}$
($      
q_{a\bu}={r^2/q_{a\ci}}={q_{\bu \ap}}=q_{\ci a}
$
).
In the limit 
$q_{a\bullet}\rightarrow 1~\forall a$, which implies  
$r\rightarrow 1$, the dilatation 
$u$ commutes with $x$ and $T$, and can be set  equal to the 
identity $I$; then,
when also  $q_{ab}\rightarrow 1$ we
recover the classical algebra $Fun(ISO(N))$.

The $\SO$ real form mentioned in the previous section
is inherited by $\ISO$.
In particular the $q$-Poincar\'e group
$ISO_{q,r}(3,1; {\mbox{\boldmath $R$}})$  
is obtained by setting $|q_{1\bu}|=|r|=1,\, q_{2\bu}/r\in 
{\mbox{\boldmath $R$}},\, q_{12}\in {\mbox{\boldmath $R$}}.$
A dilatation-free $q$-Poincar\'e group is found after the furter 
restriction $q_{1\bu}=q_{2\bu}=r=1$. The only free parameter remaining 
is then
$q_{12}\in {\mbox{\boldmath $R$}}.$  
\sk
\noi {\sl Note 1 . }
The $u$ and $x^a$ elements generate a subalgebra of
$\ISO$ because their commutation relations  
do not involve the $\T{a}{b}$ elements.
Moreover these elements can be ordered using 
(\ref{qplane}) and (\ref{ux}), and the Poincar\'e series of this 
subalgebra is the same as that 
of the commutative algebra in $N+1$ 
indeterminates \cite{FRT}.
A linear basis 
of this subalgebra is  therefore given by the ordered monomials:
$\zeta^i=u^{i_{\ci}} (x^1)^{i_{1}}...\,
(x^N)^{i_{N}}$.
Then, using (\ref{315a}) and (\ref{uT}),
a generic element of $\ISO$
can be written as $\zeta^ia_i$ where $a_i\in SO_{q,r}(N)$ and 
we conclude that $\ISO$ is a right $SO_{q,r}(N)$--module 
generated by the ordered 
monomials $\zeta^i.$

One can show that as in the classical case the expressions 
$\zeta^ia_i$
are unique: 
$\zeta^ia_i=0 \Rightarrow a_i=0 \;\forall\;\sma{$i$}$, i.e.
that $\ISO$ is a {\sl free} right $SO_{q,r}(N)$--module. 
[To prove this assertion one can show that 
(when  $q_{a\bu}=r\;\forall a$)
$\ISO$ is a bicovariant bimodule over $SO_{q,r}(N)$.
Since any bicovariant bimodule is free \cite{Wor} one 
then deduce that, 
as a right module, $\ISO$
is freely generated by the $\zeta^i$].
Similarly, writing  $a_i\zeta^i$
instead of $\zeta^ia_i$, we have that $\ISO$ is the free left 
$SO_{q,r}(N)$--module generated by the 
$\zeta^i$.
\sk
	
\sect{Universal enveloping algebra \mbox{$\U$}}
Classically the  universal 
enveloping algebra $U(g)$ of a Lie group $G$ is
the associative algebra generated by the Lie algebra $g$ of $G$. 
Given a basis $\{\chi_i\}$ of $g$, any element of 
$U(g)$ is a finite formal power 
series in the elements $\chi_i$. 
Because of the Lie algebra relations, 
it is always possible to order any product of $\chi_i$'s;
moreover 
(Poincar\'e-Birkhoff-Witt theorem)
a linear basis of $U(g)$ is given by ordered products of these elements
(for example 
$\chi_{i_1}\chi_{i_2}...\chi_{i_s}\,, i_1\leq i_2 ...\leq i_s\,,s\geq 1$).
Any element ${\chi_i}$ can be seen as a tangent vector to the 
origin of the group, and therefore it is a functional that 
associates the complex number ${\chi_i}(f)$
to any function $f$ on the group. 
In the noncommutative case, for semi-simple $g$, there is a unique 
$U_q(g)$,
while there is not a unique (categorical) definition of $q$-Lie algebra.
In this section and the next we study the universal enveloping algebras 
$\U$
and   $U_{q,r}(iso(N))$, while in Section 6 we will briefly consider 
a subspace ($q$-Lie algebra)  of $\IU$ that defines a
bicovariant differential 
calculus on
$ISO_{q,r=1}(N)$.
\sk
$\U$
is the algebra over $\mbox{\boldmath $C$}$ generated
by the counit $\epsi$ and by the functionals $\LLpm $ 
defined 
by their value on the matrix elements $\T{A}{B}$  :
\eq
\Lpm{A}{B} (\T{C}{D})= \Rpm{AC}{\!\!BD}\,, \label{LonT}
{}\Lpm{A}{B} (I)=\de^A_B 
\en
\noi with
\eq
\Rp{AC}{\!\!BD} \equiv \R{CA}{\!\!DB}\,,~
\Rm{AC}{\!\!BD} \equiv \Rinv{AC}{\!\!BD}~.
\nonumber
\en
To extend the definition (\ref{LonT}) 
to the whole algebra $\SO$ we set $\forall \,a,b\,$:
\eq
\Lpm{A}{B} (ab)=\Lpm{A}{C} (a) \Lpm{C}{B} (b)~.
\label{Lab}
\en
{}From (\ref{LonT}) 
using the upper and lower
triangularity of $R^+$ and $R^-$ we see that
$L^+$ is upper
triangular and $L^-$ is lower triangular. 
\sk
The commutations between $\Lpm{A}{B}$ 
and $\Lpm{C}{D}$ are induced by (\ref{QYB}) :
\eq
R_{12} \LLpm_2 \LLpm_1=\LLpm_1 \LLpm_2 R_{12} \label{RLL}~,
\en
\eq 
R_{12} \LLp_2 \LLm_1=\LLm_1 \LLp_2 R_{12}~, \label{RLpLm}
\en
\noi where the product $\LLpm_2 \LLpm_1$ 
is the convolution 
product $\LLpm_2 \LLpm_1 \equiv (\LLpm_2 \otimes \LLpm_1)\D$. 
\sk
The $\Lpm{A}{B}$ elements satisfy orthogonality conditions
analogous to (\ref{Torthogonality}):
\eqa
& &C^{AB} \Lpm{C}{B} \Lpm{D}{A} = C^{DC} \epsi\label{CLL1}\\
& &C_{AB} \Lpm{B}{C} \Lpm{A}{D} = C_{DC} \epsi \label{CLL2}
\ena
The co-structures of the algebra generated by the 
functionals $L^{\pm}$ and $\epsi$ 
are defined by duality [see (\ref{Lab})]:
$
\D '(\Lpm{A}{B})(a \otimes b)=\Lpm{A}{B}
(ab)\,,
\ep (\Lpm{A}{B})=\Lpm{A}{B} (I)$ 
and $\kp(\Lpm{A}{B})(a)=\Lpm{A}{B} (\kappa (a))$
 so that
\eqa
\!\!\!\!\!\!\!\!\!
& & \D ' (\Lpm{A}{B})=\Lpm{A}{G} \otimes \Lpm{G}{B}\label{copLpm}\\
\!\!\!\!\!\!\!\!\!
& & \epsi ' (\Lpm{A}{B})=\de^A_B \label{couLpm}\\
\!\!\!\!\!\!\!\!\!
& &\kp (\Lpm{A}{B})
= [(\LLpm)^{-1}]^A{}_{\!B}
= C^{DA} \Lpm{C}{D} C_{BC}
\label{coiLpm}
\ena

{}From (\ref{coiLpm}) we have that $\kp$ is an inner operation 
in the algebra generated by the
functionals $\Lpm{A}{B}$ and $\epsi$, it is then easy to see that these
elements generate a Hopf algebra, the universal enveloping algebra 
$\U$
of $\SO$.
\sk
{\sl Note 2 :} 
{}From the $CLL$ relations we have
\eq
\Lpm{A}{A}\Lpm{A'}{A'}=\Lpm{A'}{A'}\Lpm{A}{A}=\epsi ~.
\en
{}From the $RLL$ relations 
we have that the subalgebra $U^0$ generated by the elements 
$\Lpm{A}{A}$ and $\epsi$ is commutative (use upper triangularity of $R$).
\sk
\indent 
{\sl Note 3 :}
$\U$ is completely characterized by the relations 
(\ref{RLL}), (\ref{RLpLm}), (\ref{CLL1}), (\ref{CLL2}), and
$\Lm{A}{A}=F(\Lp{A}{A})$.
The functional dependence $F$ of the $\Lm{A}{A}$ in terms of the
$\Lp{A}{A}$ is studied in \cite{ISODUAL}; in the uniparametric case 
($q_{AB}\rightarrow r$) it becomes the simple expression:
\eq
\Lp{A}{A}\Lm{A}{A}
=\epsi \label{epsiaepsi}~, \mbox{ i.e.}~ \Lm{A}{A}=\Lp{A'}{A'} 
\en
[indeed $\Lp{A}{A}\Lm{A}{A}(a)=
\epsi(a)$ as can be easily seen when $a=\T{A}{B}$ 
and generalized to any $a\in \SO$ using (\ref{Lab})].
In both cases the commutative subalgebra $U^0$ can be generated
by the following elements:
$\Lp{\ci}{\ci},$  
$\Lp{1}{1}...\Lp{n}{n}$ where
$N=2n$ for $N$ even and $N=2n+1$
for $N$ odd.
\sk
The elements $\LLpm$ [or ${1\over{r-r^{-1}}}(\Lpm{A}{B}-
\de^A_B\epsi$)] 
may be seen as the quantum analogue of the tangent vectors; then
the $RLL$ relations are the quantum analogue of the Lie algebra
relations, and we can use the orthogonal $CLL$ conditions
to reduce the number of the $\LLpm$ generators to $(N+2)(N+1)/2$, 
i.e. the dimension of the classical group manifold. 

We have already given a reduced set of generators
in the case of the diagonal 
$\LLpm{}^A{}_A$. For the off-diagonal
elements, 
for example from  relation (\ref{CLL2})
choosing the free indices $(C,D)=(\bu,d)$ and using
$\Lp{\ci}{\ci}\Lp{\bu}{\bu}=\epsi$, we have:
\eq
\Lp{\ci}{d}=
-(C_{\bu \ci})^{-1}\Lp{\ci}{\ci}\,C_{ab}\Lp{b}{\bu}\Lp{a}{d}~.
\nonumber
\en
Similar results hold for $\Lp{\ci}{\bu}$, $\Lm{\bu}{a}$
and $\Lm{\bu}{\ci}$.
Iterating this procedure, from 
$C_{ab} \Lp{b}{c} \Lp{a}{d} = C_{dc} \epsi$
we find that $\Lp{1}{i}$ (with $i=2,...N-1$) and $\Lp{1}{N}$ are 
functionally dependent on $\Lp{i}{N}$ and $\Lp{1}{1}$. Similarly
for $\Lm{N}{i}$ and $\Lm{N}{1}$. The final result is that the elements
$~\Lp{a}{J}$ with $\sma{$J'<a<J$}$ and  
$~\Lm{a}{J}$ with $\sma{$J<a<J'$}$,  
whose number in both $\pm$ cases is ${1\over 4}N(N+2)$ for $N$ even 
and ${1\over 4}(N+1)^2$ for $N$ odd, 
and the elements
$\Lp{\ci}{\ci},\Lp{1}{1},...\Lp{n}{n}$
generate all $\U$. The total number of  generators is therefore 
$(N+2)(N+1)/2$.
\sk
It is also possible to show  that, as in the classical case,
these $(N+2)(N+1)/2$ generators can always be ordered.
The proof \cite{ISODUAL} is based on the observation that the full set of
$\LLpm$ can be ordered [this is shown 
by induction on $N$, using that
$U_{q,r}(so(N))$ is a subalgebra of $\U$] and that
\eq
\PA{ab}{cd} \Lp{d}{\bu} \Lp{c}{\bu}=0   
{}\;,~\PA{ab}{cd} \Lm{d}{\ci} \Lm{c}{\ci}=0   
\label{PAJ'J}
\en
[compare with (\ref{qplane})].
The conclusion is that using the  set of $(N+2)(N+1)/2$ generators 
\eq
\Lp{a}{J} \mbox{ with }\sma{$J'<a<J$}~,~ 
\Lm{a}{J} \mbox{ with } \sma{$J<a<J'$} 
\nonumber
\en
and $\Lp{\ci}{\ci},\Lp{1}{1},...\Lp{n}{n}\;,$
a generic element of $\U$ can always be written 
as a linear combination of ordered monomials of the kind
\eq
\Pi(\Lp{\al}{J})
(\Lp{\ci}{\ci})^{i_{\ci}}\!(\Lp{1}{1})^{i_1}...(\Lp{n}{n})^{i_n}
\Pi(\Lm{\al}{J})
\label{Starrr}
\en
where
$\Pi(\Lp{\al}{J})$, $[\Pi(\Lm{\al}{J})]$ is
a monomial in the off-diagonal elements $\Lp{\al}{J}$ with 
$\sma{$J'<\al<J$}$
[$\Lp{\al}{J}$ with $\sma{$J<\al<J'$}$] where an ordering has been chosen.

We conjecture that monomials of the type  (\ref{Starrr}) are linearly 
independent and therefore that the set of all these monomials is a basis 
of 
 ${U}_{q,r}(so(N+2))\,.$

\sect{Universal enveloping algebra  $U_{q,r}(iso(N))$}
Consider a generic functional $f\in\U$. It is well defined on the  
quotient
$\ISO =\SO /H$ if and only if $f(H)=0$.
It is  easy to see that the set $\H$ of all these functionals is a  
subalgebra
of $\U$:
if $f(H)=0$ and $g (H)=0$, then $fg(H)=0$ because
$\Delta  (H) \subseteq
H \otimes \SqrNt + \SqrNt \otimes H.$
Moreover $\H$ is a Hopf subalgebra of $\U$  since $H$ is a Hopf ideal
\cite{Sweedler}.
In agreement with these observations we will find the Hopf algebra
$\IU$ [dually paired to $\ISO$] as a subalgebra of $\U$ vanishing
on the ideal $H$.

Let
\eq
IU\equiv [L^{-A}{}_B, L^{+a}{}_b, L^{+\circ}{}_{\circ},
L^{+\bullet}{}_{\bullet}, \epsi]
\label{IU}
\en
be the subalgebra of $\U$ generated by
$L^{-A}{}_B, L^{+a}{}_b, L^{+\circ}{}_{\circ},   
L^{+\bullet}{}_{\bullet}, \epsi
.$
\sk
{\sl Note 4 }:  These are all and only the functionals   
annihilating
the generators of $H$:
$\T{a}{\circ}$, $\T{\bullet}{b}$ and  $\T{\bullet}{\circ}\;$.
The remaining $\U$ generators
$L^{+\circ}{}_b~,~L^{+a}{}_{\bullet}~,~L^{+\circ}{}_{\bullet}$ do not
annihilate
the generators of $H$ and are not included in
(\ref{IU}).
\sk
Since $\Delta(IU)\subseteq IU\otimes IU$ and $\kp(IU)\subseteq IU$
(as can be immediately seen at the generators level) we have that $IU$ is
a Hopf subalgebra of $\U$. 
Moreover one can also give the following $R$-marix formulation 
\cite{ISODUAL}:
\sk
\noi {\sl Theorem 5.1 }
The Hopf algebra $IU$ is generated by the unit $\epsi$ and
the matrix entries:
$$
L^-=\left(L^{-A}{}_{B_{{}_{}}}\right)
{}\,,~
\Lc=\Mat{L^{+\circ}{}_{\circ}}{}{}{}{L^{+a}{}_b}{}{}{}
{L^{+\bullet}{}_{\bullet}}
$$
these functionals satisfy the $q$-commutation relations:
\[
R_{12} \Lc_2 \Lc_1=\Lc_1 \Lc_2 R_{12} ~\mbox{ or equivalently }
\nonumber
\]
\eq 
{\cal{R}}_{12}
\Lc_2 \Lc_1=\Lc_1 \Lc_2 {\cal{R}}_{12} \label{iRLcLc}
\en
\eq
R_{12} \LLm_2 \LLm_1=\LLm_1 \LLm_2 R_{12}~, \label{iRLL}
\en
\eq
{\cal{R}}_{12} \Lc_2 \LLm_1=\LLm_1 \Lc_2 {\cal{R}}_{12}~,  
\label{iRLpLm}
\en
where
$
{\cal{R}}_{12}\equiv\Lc_2({{t}}_1)$
that is
$
{\cal{R}}^{ab}_{cd}={R}^{ab}_{cd}\,,~
{\cal{R}}^{AB}_{AB}={R}^{AB}_{AB}$
and otherwise  
${\cal{R}}^{AB}_{CD}
=0~.$ 

The orthogonality conditions and the costructures are the same 
as in 
(\ref{CLL1})--(\ref{coiLpm}), with the $\LLp$ matrix
replaced by the $\Lc$ matrix.
\sk
{\cvd}

{\sl Note 5 : } From  (\ref{iRLcLc}) applied  
to ${{t}}$ we obtain the quantum Yang-Baxter equation for the matrix  
${\cal{R}}$.
\sk
{}{\sl Note 6 : }Following {\sl Note 1} and using (\ref{PAJ'J}), 
a generic element of $IU$
can be written as $\eta^ia_i$ ($a_i\eta^i$) where $a_i\in \U$ 
and $\eta^i$ are ordered 
monomials in the $\Lm{\ci}{\ci}$ and $\Lm{a}{\ci}$  elements:
$\eta^i=(\Lm{\ci}{\ci})^{i_{\ci}}(\Lm{1}{\ci})^{i_{1}}...\,
(\Lm{N}{\ci})^{i_{N}}$. Moreover
$IU$ is a free right (left)  $U_{q,r}(so(N))$--module 
generated by the ordered 
monomials $\eta^i.$
\sk
\vskip .2cm
\noi {{{\bf{Duality}} $\IU\leftrightarrow \ISO$}}
\sk
We now show that $IU$ is dually paired to $\SO$. This is the  
fundamental
step allowing to interpret $IU$ as the universal 
enveloping  algebra 
of $\ISO$.
\sk
\noi {\sl Theorem 5.2} $~IU$ annihilates $H$, that is $IU\subseteq  
\H$.

\noi {\sl Proof :} $~~~$
Let ${\cal{L}}$ and $\Tc$ be generic generators of $IU$ and  $H$  
respectively.
As discussed in {\sl Note 4}, ${\cal{L}}(\Tc)=0$. A generic element  
of the
ideal is
given by $a\Tc b$ where sum of polynomials is understood;
we have (using Sweedler's notation for the coproduct):
\eq
{\cal{L}}(a\Tc  
b)={\cal{L}}_{(1)}(a){\cal{L}}_{(2)}(\Tc){\cal{L}}_{(3)}(b)=0~,
\en
\noi since ${\cal{L}}_{(2)}(\Tc)=0$ because ${\cal{L}}_{(2)}$ is  
still
a generator of $IU$, indeed $IU$ is a sub-coalgebra of $\U.$
Thus ${\cal{L}}(H)=0$. Recalling that a product of functionals  
annihilating
$H$ still annihilates the co-ideal $H$, we also have $IU(H)=0$.
\sk
{\cvd}
In virtue of {\sl Theorem 5.2} the following bracket is well defined:
\eqa \mbox{{\sl Definition}. }&\!\!\!\!\!\!\!\!&
\le ~ ,~ \re ~ :~ IU\otimes \ISO \longrightarrow
\mbox{\boldmath $C$}\nonumber\\
&\!\!\!\!\!\!\!\! &\le a', P(a)\re\equiv a'(a)\label{duality}\\
&\!\!\!\!\!\!\!\! &\forall \/a'\in IU ~,~\forall \/a\in \SO\nonumber
\ena
where $P~:~\SO\rightarrow \SO/H$ $ \equiv \ISO$ is the canonical
projection, which is surjective.
The bracket is well
defined because two generic counterimages of $P(a)$ differ
by an addend belonging to $H$.
\sk
Since 
$IU$ is a Hopf subalgebra
of $\U$ and $P$ is compatible with the structures and costructures
of $\SO$ and $\ISO$,
the following theorem is then easily shown, 
\sk
\noi { \sl Theorem 5.3 $~~~$} The bracket (\ref{duality})
defines a pairing between $IU$ and $\ISO$ :
$\forall\/ a',b'\in IU~,~\forall\/P(a),P(b)\in \ISO$
\eqa
& &\le a'b' , P(a)\re = \le a'\otimes b',\Delta  
(P(a))\re\label{uuno}\nonumber\\
& &\le a',P(a)P(b)\re=\le\Delta '(a'),P(a)\otimes  
P(b)\re\label{udue}\nonumber\\
& &\le\kp(a'),P(a)\re=\le a',\kappa(P(a))\re\label{utre}\nonumber\\
& &\le I,P(a)\re=\epsi(a)~;~~\le a',P(I)\re=\epsi  
'(a')\nonumber
\ena
\cvd
We now recall that $IU$ and $\ISO$, besides being dually paired, 
are free right modules respectively on $U_{q,r}(so(N))$ 
and  on $SO_{q,r}(N)$. 
They are  freely 
generated by the two isomorphic sets of the ordered monomials in the 
$q$-plane plus dilatation coordinates $\Lm{\ci}{\ci},\;\Lm{a}{\ci}$ 
and $u,\;x^a$ respectively. We then conclude that
$IU$
is the universal enveloping algebra of  $\ISO$:
$~\IU\equiv IU~.$

\sect{Projected differential calculus}
\sk
In the previous sections
we have found the inhomogeneous
quantum group  $\ISO$ by means of a projection from
$\SO$.
Dually, its universal
enveloping algebra is a given  Hopf subalgebra of
$\U$.
Using the same techniques a differential calculus on
$\ISO$ can be found.
\sk
Classically the differential calculus on a group is uniquely
determined by the Lie algebra $g$ of the tangent vectors to the group.
Similarly a differential calculus on a $q$-group $A$, with universal 
enveloping algebra ${\cal{U}}$ (${\cal{U}}\equiv U_q(g)$), is determined 
by 
a 
$q$-Lie algebra $T$: the $q$-deformation of $g$. It is natural
to look for  a linear space $T$ , $T\subset \mbox{ker}\epsi\subset 
{\cal{U}}$
satisfying the following three conditions:
\eq
T ~\mbox{ generates }~{\cal{U}}\label{ZERO}
\en
\eq
\D '(T)\subset T\otimes \epsi + {\cal{U}}\otimes T ,\label{UNO}
\en
\eq
[T,T]\subset T \label{DUE}
\en
where the braket is the adjoint action defined by
\eq
\forall\, 
\chi, \psi \in {T}~~[\chi,\psi]=\chi_1\psi\kappa'(\chi_2)~.
\en
Condition (\ref{UNO}) states that 
the  elements of $T$ are generalized tangent vectors, 
and in fact, if  $\{\chi_i\}$ is a basis of the linear space $T$, 
we have 
$
\D(\chi_i)=\chi_i\otimes \epsi +f_i{}^j \otimes \chi_j\label{leibdef}
$ that is equivalent to   
\eq
\chi_i(ab)=\chi_i(a)\,\epsi(b) + f_i{}^j(a)\,\chi_j(b)
\en
where  $f_i{}^j\in {\cal{U}}$ and $\epsi(f_i{}^j)=\delta_i^j$.
[Hint: apply  
$(\epsi\otimes id)$  to (\ref{UNO})]. 
In the commutative limit we espect $f_i{}^j\rightarrow 1$.
One can also consider the generalized left-invariant 
vector fields $\chi_i*$ defined by 
$\chi_i*a\equiv a_1\chi_i(a_2)$, 
then (\ref{UNO}) states that the $\chi_i*$ are generalized 
derivations:
\eq
\chi_i*(ab)=(\chi_i*a)\,b + (f_i{}^j*a)\,\chi_j*b~.
\en
Condition (\ref{DUE}) is the closure of 
$T$ under the adjoint action,
in the classical case, if $\chi$ is a tangent vector: $\D(\chi)
=\chi\otimes \epsi +\epsi\otimes \chi\,,~ \kappa'(\chi)=-\chi$
and the adjoint action of $\chi$ on $\psi$ is given by the 
commutator $\chi\psi -\psi\chi$.
\sk
\noi {\sl Note 7 :}
Woronowicz conditions for a bicovariant differential
calculus are slightly weaker than (\ref{ZERO})-(\ref{DUE}). 
Relation (\ref{UNO}) can also be written
$\D (T\oplus\{\epsi\})\subset {\cal{U}}\otimes  (T\oplus\{\epsi\}) $,
where $T\oplus\{\epsi\}$ is the vector space spanned by $\chi_i$ and 
$\epsi$;
therefore $\kp(T\oplus\{\epsi\})$ is a right {\sl co-ideal}, 
it is the space orthogonal to the Woronowicz \cite{Wor} right {\sl ideal} 
$R$:
$[\kp(T\oplus\{\epsi\})](R)=0$.
Relations (\ref{ZERO}) and (\ref{DUE}) imply that
$\forall\varphi\in {\cal{U}}, \forall\chi\in T, ~
\varphi_1\chi\kappa'(\varphi_2)\in T$
(for example if $\varphi=\chi''\chi'$ then
$\varphi_1\chi\kp(\varphi_2)=[\chi'',[\chi',\chi]])$;
this last condition
is then equivalent to the $ad$ invariance of $R$:
$ad(R)\subset R\otimes A$, where $ad(a)\equiv a_2\otimes \kappa(a_1)a_3$.
{}For further details see \cite{SWZ}.
\sk
{}From  (\ref{ZERO})--(\ref{DUE}),
the construction of the differential calculus associated 
to the tangent space $T$
is quite straighforward (see for example 
\cite{PaoloPeter}): the main ingredients are

{\bf{i) }} the coordinates $x^j$, that are uniquely defined by 
$x^j\in$ ker$\epsi - \kappa(R),~\chi_i(x^j)=\delta^j_i$;

{\bf{ii) }} the adjoint representation 
$M_i{}^j\equiv\chi_i(x_2^j)x_3^j\kappa^{-1}(x_1^j)$,
that satisfies $\D({M_i}^j)=M_i{}^k
\otimes M_k{}^j$ and $\epsi(M_i{}^j)=\delta^i_j$; 
note that if $y^j\in$ ker$\epsi$ and $\chi_i(y^j)=\delta^j_i$ then
$M_i{}^j=\chi_i(y_2^j)y_3^j\kappa^{-1}(y_1^j)$.

{\bf{iii) }} the space of left invariant one-forms, 
defined as the space dual to that of the
tangent vectors: $~\omega^i:\,\le \chi_i~,~\omega^j\re=\delta^j_i$.
A generic one-form is then given by
$\rho=\omega^ia_i$.
[The space of one-forms is the bicovariant bimodule
freely generated by the $\omega^i$ with 
$a\omega^i=\omega^jf_j{}^i*a\equiv
\omega^j(id\otimes f_j{}^i)\D(a),~ $ $\D{}_L \omega^i\equiv I\otimes 
\omega^i,~ \D{}_R \omega^i\equiv\omega^j\otimes M_j{}^i$].

{\bf{iv) }} The differential, defined by $da=\omega^i\chi_i*a$; it
satisfies the undeformed Leibniz rule.
\sk
In our case $A=\SO$, and a differential calculus satisfying 
(\ref{ZERO})--(\ref{DUE})
is given by the $\chi$ functionals [see \cite{Burr} for (\ref{ZERO})]:
\eq
\cchi{A}{B} = {-1\over{r-r^{-1}}} [f^{A}_{~B C}{}^{\!C}-\de^A_B \epsi]~~;
\en
the $f$ functionals and the adjoint representation read
\eq
f^{A_1}_{~A_2B_1}{}^{\!B_2} \equiv {\kp}^{-1} (\Lm{B_2}{A_2}) 
\Lp{A_1}{B_1},
\label{defff}
\en
\eq
\MM{A}{BC}{D}\equiv\T{A}{C}\kappa(\T{D}{B})~;\label{last?}
\en
see  \cite{Jurco} and references therein (see also  
\cite{altroarticolo}, and for notations \cite{PaoloPeter}, 
to obtain the $\chi_i$ of \cite{altroarticolo} act on our $\chi_i$ 
with $-\kp$).
\sk
We now consider the differential calculus on $\ISO$.
{}From (\ref{UNO}) and (\ref{DUE}) it is immediate to see that 
$T'\equiv T\cap \IU$ satisfies $\D(T')\subset T'\otimes\epsi+
\IU\otimes T'$ and  $[T',T']\subseteq T\cap\IU=T'$;
indeed $\IU$ is a Hopf subalgebra of $\U$. 
However, condition (\ref{ZERO})
is not fulfilled since the $\cchi{a}{b}$ do not belong to $\IU$
and therefore to $T'$ unless $r=1$. [The $\cchi{a}{b}$ 
 contain the addend ${-1\over r-r^{-1}}f^{a}_{~b\bu}{}^{\!\bu}\not\in 
\IU$, 
that vanish  for $r\rightarrow 1$]. 
We therefore obtain an $ISO_{q,r=1}(N)$ bicovariant differential
calculus. 
A basis of $T'$ is given by 
\eq
\cchi{a}{b}=\limrone\linv [f^{a}_{~b c}{}^{\!c}-\de^a_b \epsi]\,,~
\mbox{with  \sma{$a+b> N+1~$}};\label{ISOtang1}
\en
\eq
\cchi{\bu}{b}=\limrone\linv f^{\bu}_{~b\bu}{}^{\!\bu}\,;~
\cchi{\bu}{\bu} = \limrone\linv [f^{\bu}_{~\bu\bu}{}^{\!\bu} - \epsi]\,,
\label{ISOtang2}
\en
here $\lambda=r^{-1}-r$; notice that all the other 
$\cchi{a}{b}$ in the $r\rightarrow 1$ limit are linearly
dependent from the $\cchi{A}{B}$ in (\ref{ISOtang1}) and (\ref{ISOtang2})
\cite{altroarticolo}.
The $q$-Lie algebra is
\eqa
\!\!\!\!\!\!\!\!\!\!\!\!& &\cchi{b_{1}}{b_2}  
\cchi{c_{1}}{c_2} - (q_{b_1c_2} q_{c_1b_1}
q_{b_2c_1} q_{c_2b_2})
{}~\cchi{c_{1}}{c_2} \cchi{b_{1}}{b_2}=~~~~~~~~~\nonumber\\
\!\!\!\!\!\!\!\!\!\!\!\!
& &~~~~+(q_{b_1c_2} q_{c_2b_2} q_{b_2b_1} \de^{c_1}_{b_2})
{}~\cchi{b_{1}}
{c_2}-
(q_{c_1b_1} q_{b_2b_1} C_{b_2c_2}) \cchi{b_{1}}{c_1'}\nonumber\\
\!\!\!\!\!\!\!\!\!\!\!\!& &~~~~-(q_{c_2b_2} q_{b_1c_2} 
C^{c_1b_1}) \cchi{b_2'} {c_2}+
(q_{b_2c_1} \de^{b_1}_{c_2})\cchi{b_2'}{c_1'}~,
\nonumber
\ena
\eqa
\!\!\!\!\!\!\!\!\!\!\!\!\!\!\!\!\!\!\!\!& &\chi_{b_2} \cchi{c_1}{c_2} -
({q_{c_1\bu} \over q_{c_2\bu}}
q_{b_2c_1} q_{c_2b_2}) \cchi{c_1}{c_2}\chi_{b_2} \nonumber\\
\!\!\!\!\!\!\!\!\!&&~~=-
{q_{c_1\bu} \over q_{c_2\bu}}[C_{b_2c_2} \chi_{c_1'}-\de^{c_1}_{b_2}
q_{c_2c_1} \chi_{c_2}]~, \nonumber\\
\!\!\!\!\!\!\!\!\!\!\!\!\!\!&&\chi_{b_2} \chi_{c_2} - 
({q_{b_2\bu} \over q_{c_2\bu}}
q_{c_2b_2}) \chi_{c_2} \chi_{b_2}
=0~,\nonumber\\
\!\!\!\!\!\!\!\!\!\!\!\!\!\!\!\!\!\!&&\cchi{\bu}{\bu}  
\cchi{c_1}{c_2} -\cchi{c_1}{c_2} \cchi{\bu}{\bu}=0 ~,~\,
\cchi{\bu}{\bu}\chi_{c_2} - \chi_{c_2}\cchi{\bu}{\bu}=\chi_{c_2}\,
\nonumber
\ena
\noi where we have defined
$\chi_a \equiv  \cchi{\bu}{a}~.$ The exterior differential reads
\eq
da=\omega_{a}{}^b (\cchi{a}{b}*a)+
\omega_{\bu}{}^b (\cchi{\bu}{b}*a)+
\omega_{\bu}{}^{\bu}(\cchi{\bu}{\bu}*a)~~;\nonumber
\en
where $\omega_{a}{}^{b},~\omega_{\bu}{}^{b},$
and $\omega_{\bu}{}^{\bu}$ are the one-forms dual
to the tangent vectors (\ref{ISOtang1}) and (\ref{ISOtang2}).
\sk
In the limit $r\rightarrow 1\,,q_{a\bu}\rightarrow 1$
we have seen from (\ref{uT}) and (\ref{ux}) that 
the element $u$
can be set equal to the identity $I$, i.e. we obtain an $\ISO$
$q$-group without dilatations.
Then the tangent vectors $\cchi{a}{b}$ and $\cchi{\bu}{b}$ 
alone generate $\IU$, they satisfy (\ref{UNO}) and (\ref{DUE}) 
and therefore we have an inhomogeneous $q$-Lie
algebra and a bicovariant differential calculus without dilatation
generator $\cchi{\bu}{\bu}$.
In particular, for $N=4$ we have a  one  deformation parameter 
($q_{12}\in \mbox{\boldmath $R$}$) differential calculus on the 
$q$-Poincar\'e group discussed at the end of Section 3.
\sk
\noi {\bf Acknowledgments}
\sk
The author thanks the organizers of the 30-th Ahrenshoop Symposium 
for their invitation and support;
he  is also grateful to P. Schupp for useful discussions.


\end{document}